# Distinguishing nonlinear processes in atomic media via orbital angular momentum transfer

Alexander M. Akulshin,[1,][*] Russell J. McLean,[1] Eugeniy E. Mikhailov,[2] and Irina Novikova[2]

[1]*Centre for Quantum and Optical Science, Swinburne University of Technology, Melbourne, Australia*
[2]*College of William & Mary, Williamsburg, VA 23185, USA*
(Dated: December 19, 2014)

We suggest a technique based on the transfer of topological charge from applied laser radiation to directional and coherent optical fields generated in ladder-type excited atomic media to identify the major processes responsible for their appearance. As an illustration, in Rb vapours we analyse transverse intensity and phase profiles of the forward-directed collimated blue and near-IR light using self-interference and astigmatic transformation techniques when either or both of two resonant laser beams carry orbital angular momentum. Our observations unambiguously demonstrate that emission at 1.37 $\mu$m is the result of a parametric four-wave mixing process involving only one of the two applied laser fields.

Interaction of laser radiation with atomic vapours gives rise to a wide range of nonlinear processes, often resulting in the generation of new optical fields through various frequency up- and down-conversion mechanisms [1]. In some cases several physical mechanisms can be involved and identification of the origin of the new field may become nontrivial. In the case of atoms excited to higher energy levels, sufficient optical gain on some of the optical transitions can give rise to both parametric wave mixing and amplified spontaneous emission (ASE). The competition between these processes followed by the emission of infrared (IR) and blue light was investigated in alkali vapours excited by intense pulsed radiation [2–4].

A number of more recent experiments have firmly established the parametric four wave mixing (FWM) origin of the forward-directed collimated blue light (CBL) generated in Rb [5–11] and Cs [12] vapours excited by low-power cw laser radiation. However, identification of processes responsible for generation of the polychromatic IR emission (Fig. 1a) is less straightforward. Recent study of the mid-IR radiation at 5.2 $\mu$m, which is an essential part of the FWM loop that results in CBL generation in the case of Rb atoms, revealed that both ASE and parametric wave mixing processes are essential for its appearance [13]. Existence of two decay channels of the $6S_{1/2}$ level results in a competition between processes responsible for the generation of the near-IR emission at 1.32 $\mu$m and 1.37 $\mu$m. At relatively low Rb density ($N \leq 6 \times 10^{12}$ cm$^{-3}$) only forward-directed collimated light at 1.37 $\mu$m is generated [13]. However, at higher $N$ collimated emission at 1.32 $\mu$m dominates, having been observed in both the co- and counter-propagating directions and attributed to non-parametric ASE on the $6S_{1/2} \rightarrow 5P_{1/2}$ transition [14].

Appearance of the CBL and near-IR emission at 1.37 $\mu$m in only the co-propagating direction suggests a common parametric origin. Thus, the collimated radiation at 1.37 $\mu$m could be a product of a FWM process that includes the applied laser field at 776 nm and two internally generated fields at 5.23 and 2.73 $\mu$m, although the possibility of the six-wave mixing (SWM) process that involves the Rb ground state ($5S_{1/2} - 5P_{3/2} - 5D_{5/2} - 6P_{3/2} - 6S_{1/2} - 5P_{3/2} - 5S_{1/2}$) cannot be ruled out, as new fields at both 1.32 $\mu$m and 1.37 $\mu$m recently observed in very dense Rb vapours ($N \geq 2.5 \times 10^{15}$ cm$^{-3}$) were attributed to parametric SWM [15].

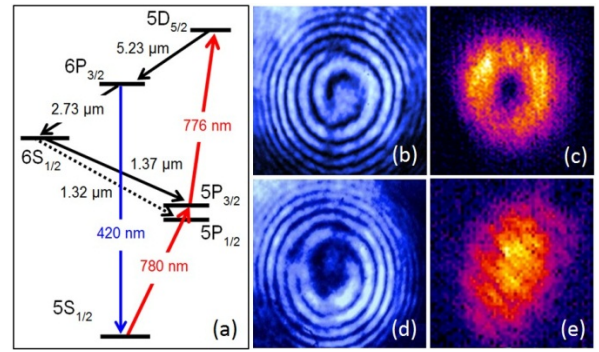

FIG. 1. (a) A diagram of the Rb atom energy levels involved in wave mixing processes and new field generation. Collimated blue light self-interference patterns (b, d) and forward-directed 1.37 $\mu$m intensity profiles (c, e) obtained with vortex laser beam at 776 nm (b, c) or at 780 nm (d, e), respectively.

In this work we investigate the origin of the forward-directed collimated near-IR emission at 1.37 $\mu$m in the low atomic density regime by analysing the transfer of orbital angular momentum (OAM) from the applied laser radiation to the frequency up- and down-converted optical fields. This idea exploits the well-known fact that new fields generated in FWM processes accumulate OAM of the applied laser light [16].

Images presented in Figure 1 show self-interference patterns of the forward-directed CBL and transverse intensity profiles of the near-IR emission obtained with the vortex laser light either at 776 nm or 780 nm. To generate forward-directed collimated emission at 420 nm and 1.37 μm in Rb vapours we use an experimental setup similar to that previously used [7, 13], and shown schematically in Fig. 2. In the self-interference method a fraction of the original vortex beam is expanded sufficiently to produce an almost flat wavefront, then mixed again with the unexpanded beam, as shown in the dashed box in Fig. 2.

[*] aakoulchine@swin.edu.au

In the case of vortex laser light at 776 nm and plane phase wavefront radiation at 780 nm the transverse intensity profile of the CBL displays the characteristic doughnut-type intensity distribution. The CBL topological charge is determined from the self-interference pattern shown in Fig. 1(b). A clearly visible single spiral structure convincingly demonstrates that the CBL has a helical wavefront, carrying the OAM $\ell = 1$ transferred from the laser light. The low intensity of the 1.37 μm radiation prevents us from analyzing it with the self-interference method, however, image (c) shows that its intensity profile also possesses a doughnut shape.

The images shown in Fig. 1(d, e) correspond to the case when Rb atoms interact with vortex light at 780 nm and plane phase-front radiation at 776 nm. As in the previous case, a high-contrast spiral structure in the interference pattern indicates that OAM has been transferred from the laser radiation via the FWM process to the CBL. However, image (e), which represents the intensity profile of near-IR radiation, does not show any intensity reduction in the centre, which is an essential feature of vortex-bearing radiation. This means that OAM has not been transferred to the 1.37 μm emission. It also suggests that the laser field at 780 nm serves only to populate the $5P_{3/2}$ level and is not involved directly in the coherent wave mixing process that results in 1.37 μm light generation. Thus, the experimental observations are consistent with our expectations based on the suggested FWM mechanism for the near-IR emission at 1.37 μm.

Image (c) of Fig. 1 cannot, however, be considered as conclusive proof of OAM transfer to the 1.37 μm light. The reason is that the transverse intensity distribution of conical emission, which is very common for optical fields generated by parametric FWM processes [7, 18, 19], can be almost identical to the doughnut-shaped intensity profiles of vortex emission. Because the 1.37 μm radiation was too weak to produce a stable spiral-type self-interference pattern to verify its vortex nature, an alternative simple method of topological charge determination [20, 21] was used. This method relies on the fact that a monochromatic optical vortex with topological charge $\ell$ splits into $|\ell|$ elementary vortices under astigmatic transformation, revealing $|\ell|$ tilted dark stripes in its image near the focus. This astigmatism is introduced simply by using a tilted $f = 1$ m lens in place of the self-interference scheme (Fig. 2).

Before applying this method to the forward-directed near-IR emission, its ability to distinguish conical and vortex emission has been first tested using CBL. Figure 3(a, b) show the vortex CBL intensity profiles recorded without and with lens tilt, respectively. Image (a) displays the familiar doughnut-shaped transverse intensity profile. In the tilted-lens case the dark stripe across the CBL image (b) indicates that a single topological charge has been transferred to the CBL from the vortex laser light at 776 nm. Image (c) shows that under certain experimental parameters the conical CBL, which carries zero OAM as it is generated by applied laser light with plane wavefronts, could have an intensity profile that is very similar to vortex-carrying CBL. However, the absence of a dark stripe across the image (d) in the tilted lens case allows their unambiguous distinction.

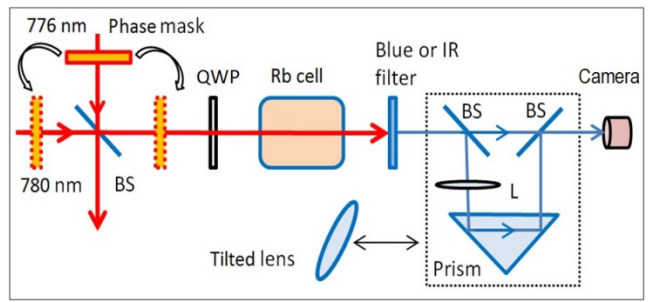

FIG. 2. Schematic of the experimental setup. Rubidium atoms are ladder-type excited to the $5D_{5/2}$ level by resonant radiation derived from two extended cavity diode lasers, tuned to the $^{85}$Rb $5S_{1/2} \rightarrow 5P_{3/2}$ and $5P_{3/2} \rightarrow 5D_{5/2}$ transitions, respectively. The optical frequency of the 780 nm laser is locked to a Doppler-free polarization resonance on the $5S_{1/2}(F = 3) \rightarrow 5P_{3/2}(F' = 4)$ transition (not shown), and the frequency of the 776 nm laser is adjusted to maximize the efficiency of frequency up- and down-conversion. The two laser beams are combined on a non-polarizing beam-splitter and converted into circularly polarized bi-chromatic radiation using a quarter-wave plate (QWP) before being focussed with a 20-cm focal-length lens into a 5-cm long cylindrical glass cell containing Rb vapour with natural isotopic abundance. Maximum laser powers at 780 and 776 nm before entering the cell are 5 and 3 mW, respectively, and can be further attenuated with variable neutral density filters. The temperature of the cell is controlled by a resistive heater, so that the atom density $N$ of saturated Rb vapour varies in the range $0.3 - 1.5 \times 10^{12}$ cm$^{-3}$. A transparent eight-octant single-charge spiral phase mask with variable azimuthal thickness is used to convert either or both pump beams into vortex beams with topological charge $|\ell| = 1$ [17]. Spatial properties of the near-IR emission are monitored using a Xeneth camera.

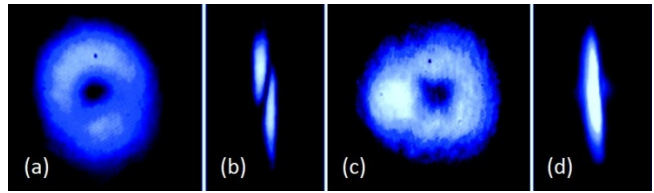

FIG. 3. Distinguishing of vortex (a, b) and conical (c, d) collimated blue light using the tilted-lens method.

A comparison of the OAM transfer from resonant laser light to the near-IR and CBL beams using the tilted-lens method is shown in Figure 4. The first two images (a, b) are obtained with vortex laser light at 780 nm and plane wavefront radiation at 776 nm. The $|\ell|=1$ topological charge of the CBL is revealed by the clear dark stripe across image (a), while the smooth image (b) indicates that the 1.37 $\mu$m beam is not vortex-bearing. However, in the case of the applied vortex laser light at 776 nm and flat phase profile at 780 nm, both CBL and the near-IR emission appear to carry the same OAM, as suggested by the similarly inclined dark stripes on the corresponding images (c) and (d). This supports our previous observations regarding OAM transfer from the laser to near-IR light. The opposite inclination of the dark stripe on the image (a) compared to the others is due to the different orientation of the phase mask in the laser beams.

The spiral structures and dark stripes observed with vortex CBL convincingly demonstrate that the OAM carried by the applied laser light either at 780 or 776 nm is transferred into the CBL, as both laser fields are involved into the FWM process. By contrast, analysing the phase wavefront of the collimated forward-directed light at 1.37 $\mu$m, we find that the transfer of the topological charge occurs from the 776 nm laser field only. If OAM is carried by the radiation at 780 nm, the phase profile of the 1.37 $\mu$m emission is indistinguishable from profiles of plane-wavefront beams.

These observations are in good agreement with the suggested parametric FWM origin of the co-propagating emission at 1.37 $\mu$m generated through the $5P_{3/2} - 5D_{5/2} - 6P_{3/2} - 6S_{1/2} - 5P_{3/2}$ cycle. Indeed, since the 776 nm laser field is a part of this loop, its non-zero OAM must be transferred to one of the emitted fields, and following the arguments presented in Ref. [10], we expect the OAM to be transferred preferentially to the generated field with spatial profile most similar to that of the OAM-carrying 776 nm, which in this case is the detected 1.37 $\mu$m field. On the other hand, as the 780 nm pump field is not directly involved in this FWM process, its OAM cannot be transferred to the forward-directed emission at 1.37 $\mu$m, as has been experimentally confirmed.

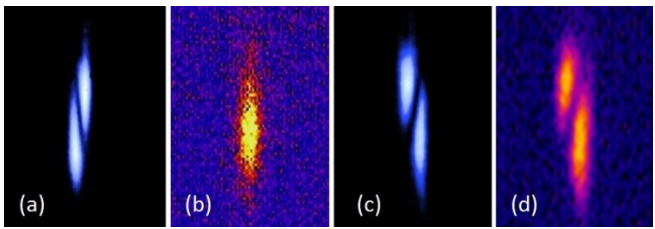

FIG. 4. Intensity profiles of the blue (a, c) and the 1.37 $\mu$m (b, d) optical beams near the focal distance of the tilted-lens. The first two images (a, b) correspond to a plane wavefront 776 nm beam and $\ell = +1$ topologically charged 780 nm beam, while the other two (c, d) are obtained with a plane wavefront 780 nm beam and $\ell = -1$ topologically charged 776 nm beam.

However, the intensity distribution of the applied laser radiation at 780 nm significantly affects the overall efficiency of the CBL and 1.37 $\mu$m light generation.

Finally we place the phase mask into the combined bichromatic laser beam, simultaneously transforming both the pump fields into vortex-carrying light with the same $|\ell|=1$ topological charge. In this case both the radiation at 420 nm and 1.37 $\mu$m display doughnut-shaped intensity profiles. The phase wavefront analysis using the self-interference and tilted-lens methods reveals that the CBL beam carries the topological charge $|\ell|=2$, indicated by the double spiral structure and two dark stripes in Figure 5(a, b). This is in good agreement with previous observations that the full topological charge carried by the pump lasers is always transferred into the CBL beam [10]. At the same time, the single dark stripe in the tilted-lens image (c) indicates that the topological charge of the 1.37 $\mu$m beam is $|\ell|=1$. Again, this observation is consistent with the suggested FWM description, as the sum of pump field OAM must be carried out by CBL, while only the topological charge at 776 nm is transferred to the collimated forward-directed IR radiation.

We note that the observed shapes of self-interference patterns and transverse intensity profiles are stable against some variation in experimental parameters. For example, the powers of the 780 nm and 776 nm lasers before entering the cell can be reduced by up to 50 % from their maximum values of 5 and 3 mW, respectively, and beam overlapping can be misaligned by a few mrad. Changing the Rb number density in the range $0.3 - 1.5 \times 10^{12}$ cm$^{-3}$ by heating the cell affects the intensities of the new fields, but not their spatial characteristics.

Our observations demonstrate the usefulness of using vortex-carrying light as a marker for identifying the processes involved in the new-field generation, in particular, considering the possibility that the near-IR field is generated by a SWM process. In principle, four-wave and six-wave mixing processes can not only co-exist in atomic media [22, 23], but they could be almost equally efficient, producing spatial and temporal interferences [24]. However, as was discussed in [10], the total angular momentum of the two pump fields is more likely to be transferred to the emitted field having the closest spatial overlap with the applied laser beams, which, in case of the six-wave mixing, would be the re-emission at the $5P_{3/2} \rightarrow 5S_{1/2}$ transition, being the most similar to both pump field modes. In this scenario we would expect to observe no OAM transfer to the 1.37 $\mu$m optical field if SWM was responsible for its generation, even when the 776 nm optical field is a vortex beam. This clearly contradicts the experimental observation. Moreover, the suggested method may be even more useful in identifying whether the 1.32 $\mu$m radiation is generated via a SWM process or as a result of ASE, since in the case of SWM the directional radiation emitted on the $5P_{1/2} \rightarrow 5S_{1/2}$ transition would be expected to carry the OAM, and because its wavelength is diffrent from either of the applied laser fields its spatial and phase profiles can be easily analyzed.

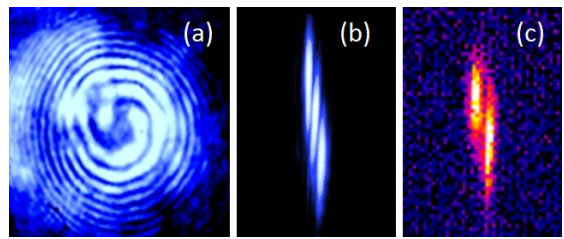

FIG. 5. Analysis of the topological charge of the blue (a, b) and near-IR (c) light generated with both vortex-bearing components of the bichromatic laser beam.

In conclusion, we have experimentally investigated the orbital angular momentum transfer from one or both of the applied laser fields at 780 and 776 nm, which pump Rb atoms to the $5D_{5/2}$ excited level, to the blue and near-IR emission generated in the co-propagating direction. We observe that the blue light is emitted with topological charge equal to the sum of the topological charges of the two pump fields, as expected for a phase-matched four-wave mixing process. At the same time, the OAM of the collimated near-IR emission at 1.37 $\mu$m is always equal to that of the 776 nm laser field, which is the only applied laser radiation that participates in the $5P_{3/2} - 5D_{5/2} - 6P_{3/2} - 6S_{1/2} - 5P_{3/2}$ four-photon loop. Thus, we believe that our experiment unambiguously demonstrates that the generated 1.37 $\mu$m optical field is the product of a parametric FWM process.

The experimental procedure based on topological charge transfer combined with the simple and robust method for orbital angular momentum measurements can be useful for distinguishing nonlinear processes in atomic media.

The authors thank Dmitry Budker and Laurence Pruvost for useful discussions, Grover Swartzlander for the loan of the phase mask and Qiming Zhang for lending the Xeneth IR camera. They also gratefully acknowledge the Swinburne University Visiting Researcher Scheme for supporting visits of Irina Novikova and Eugeniy Mikhailov making this collaboration possible. IN and EEM also acknowledge the support of AFOSR grant FA9550-13-1-0098 and NSF grant PHY-1308281.